\documentclass[]{spie}  

\usepackage{amsmath,amsfonts,amssymb}
\usepackage{graphicx}
\usepackage[colorlinks=true, allcolors=blue]{hyperref}

\newcommand{\aap}{Astronomy and Astrophysics}
\newcommand{\apj}{The Astrophysical Journal}

\newcommand{\pasp}{Publications of the Astronomical Society of the Pacific}
\newcommand{\procspie}{Proceedings of the SPIE}
\newcommand{\apjl}{The Astrophysical Journal Letters}
\newcommand{\aj}{The Astronomical Journal}

\title{ESCAPE project \\ CAPyBARA: a Roman Coronagraph simulator for post-processing methods development}

\author[a]{Lisa Altinier}
\author[a]{Élodie Choquet}
\author[a]{Arthur Vigan}
\author[a]{Nicolás Godoy}
\author[a]{Alexis Lau}
\affil[a]{Aix Marseille Univ, CNRS, CNES, LAM, Marseille, France}

\begin{document} 
\maketitle

\begin{abstract}

The Roman Coronagraph Instrument will be the first space facility equipped with deformable mirrors (DMs). These will lead to reach a contrast of $10^{-8}$ or better in a dark hole between $3 \lambda / D$ and $9 \lambda / D$. Post-processing techniques play an important role in increasing the contrast limits. 

Our work investigates how DMs can be used to calibrate the instrument response to controlled wavefront error maps and to improve the post-processing performance. To this goal, we are developing a simulation pipeline, CAPyBARA, that includes both a propagation model of the Coronagraph and a post-processing module and produces starlight subtracted images of a science target. This pipeline will then allow us to investigate alternative observing strategies and test their performance for the Roman Coronagraph.

Here we present the first version of the simulator: it currently reproduces the optical propagation, which consists in the hybrid Lyot coronagraph (HLC) optical structure and dark-hole digging technique (Electric Field Conjugation, EFC, coupled with $\beta$-bumping), the environment (quasi-static aberration) and the post-processing.
With it, we mimic a Roman Coronagraph Instrument observing sequence, which consists in first acquiring reference star data before slewing to the scientific target, and we investigate how the evolution of the quasi-static aberrations deteriorate the contrast limit in the dark hole. We simulate a science target with planets at high contrast with their star and we perform a first post-processing analysis with classical subtraction techniques. 
Here we present the CAPyBARA simulator, as well as some first results.
The next step will be to generate PSF libraries by injecting pre-calibrated probes on the DMs (in open loop) during the reference star acquisition and compute a PCA model. Later, we will compare the performance gain obtained with the modulated-DM reference library over standard approaches (RDI).

\end{abstract}


\section{INTRODUCTION}
\label{sec:intro}  

The Nancy Grace Roman Space Telescope (hereafter Roman) is the next NASA space telescope. Selected during the 2010 Decadal survey, it is a Hubble-class telescope equipped with a monolithic primary mirror of $2.4 \ m$ working in the visible and near-infrared. The launch is foreseen for late 2026 - early 2027, and it will set an important milestone in terms of technological demonstration for the upcoming Habitable Worlds Observatory (HWO): one of its instruments, the Coronagraph Instrument, will be equipped, for the first time in space, with deformable mirrors (DMs), which will be fundamental for the improvement of the detection limit of exoplanets in direct imaging mode. The contrast requirement for the technological demonstration to be successful is $10^{-7}$ in an area between $6 - 9 \ \lambda / D$ in Band 1 (i.e. $575 \ \texttt{nm}$ with $10 \%$ of bandwidth) \cite{mennesson2021_demotech, bailey2022_demotech, Bailey2023}, but the Coronagraph Instrument, and in particular the hybrid Lyot coronagraph (HLC) is designed and built to reach up to $\sim10^{-9}$ after post-processing \cite{Bailey2023}. After the launch, $90$ days will be dedicated to the commissioning \cite{bailey2022_demotech,mennesson2021_demotech}, technological demonstration and scientific observations.

\begin{figure} [ht]
\begin{center}
\begin{tabular}{c} 
\includegraphics[height=8cm]{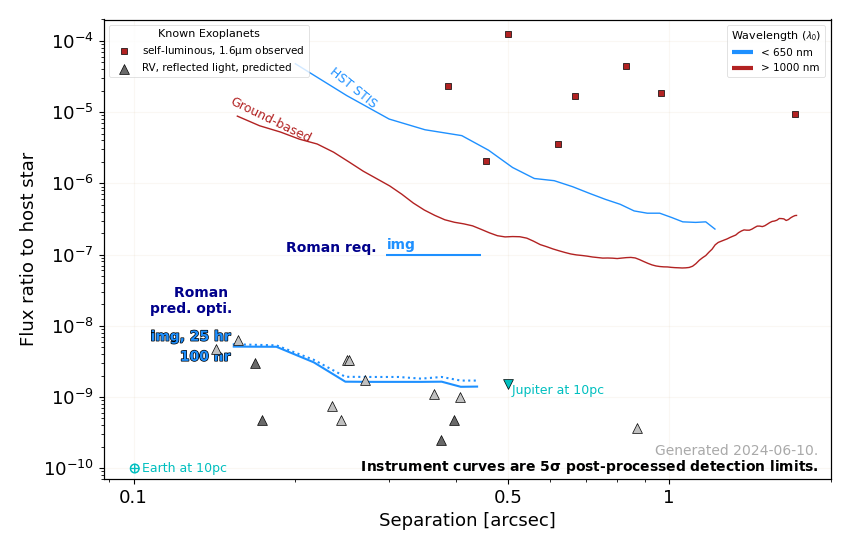}
\end{tabular}
\end{center}
\caption[example] { \label{fig:plot_contrast_comparison_rst} Comparison of the flux ratio and separation from the host star between detected self-luminous exoplanets (red squares) and the known jovian exoplanets detected with radial velocity method and never imaged before (gray triangles). These last, will be possible to be imaged in reflected light by Roman. As a reference, instrumental limits are plotted. Plot adapted from \citenum{Bailey2023}}
\end{figure}

Planets imaged up to now are young (less than $500 \text{Myr}$ \cite{Chauvin2018}) and thus still self-luminous. Moreover, they orbit relatively far from their host star (beyond $10 \ \texttt{au}$ \cite{Bowler2016, Nielsen2019, Vigan2021}) and their contrast ratio is at best $10^{-6}$ (Figure \ref{fig:plot_contrast_comparison_rst}). Roman Coronagraph will explore an area surrounding the star, between $0.15$ and $0.44 \ \text{arcsec}$, that has never been probed in direct imaging with such high contrasts. This is possible thanks to the technique used for the dark hole digging that is enabled by the DMs in the instrument, the Electric Field Conjugation (EFC \cite{giveon_2007, giveon_2007_b, potier2020, Krist2023_the_bible, Zhou2023}). With it, in Roman observational bandwidth ($0.48 - 2.3 \ \mu \text{m}$), the contrast ratio will be improved by three order of magnitudes. This will allow, for the first time, the observation of giant planets in reflected light, which are older and not self-luminous. 
Post-processing techniques play an important role in pushing these limits toward higher contrast ratio. However, current methods are either optimised for ground-based facility equipped with adaptive optics (hereafter AO) or for space-based telescopes without wavefront control. 

In this context, the ESCAPE project (Exoplanet Systems with a Coronagraphic Active Processing Engine, Choquet et al, 2024 \cite{Choquet2024}, these proceedings) has the goal of investigating the potential of active components in space to improve post-processing performance, to improve the detection of exoplanets with high-contrast imaging. Roman represents a unique opportunity to demonstrate these methods on sky.

\begin{figure} [ht]
\begin{center}
\begin{tabular}{c} 
\includegraphics[height=8cm]{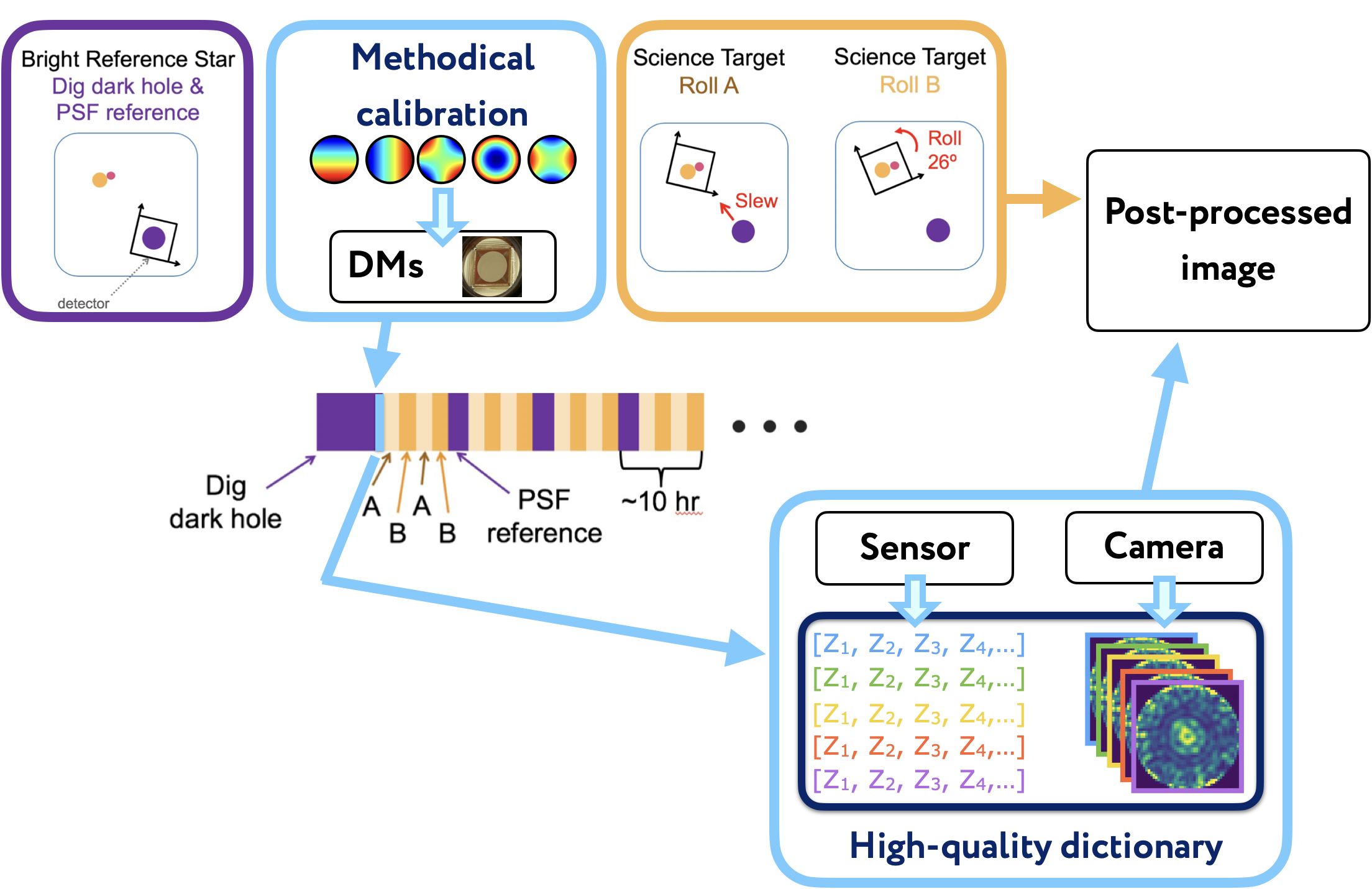}
\end{tabular}
\end{center}
\caption[example] { \label{fig:escape_lisa} Roman observing sequence \cite{Bailey2023}, where we inserted the Zernike calibration maps we want to test on Roman. The output of the calibration will be registered both for the sensor and for the camera, in a high contrast dictionary. This catalogue will be used for post-processing.}
\end{figure}

As part of the ESCAPE project, we are developing a package to simulate Roman coronagraphic observations, including all the steps from simulating images with HLC, to the post-processing analysis. This will be used to inject controlled aberration maps on the DMs. By recording the wavefront sensor telemetry and on the acquired images on the camera, we will be able to build a high-quality dictionary that can be used during the post-processing analysis. The method will be tested by injecting synthetic planets in the mock data. The procedure described is summarised in Figure \ref{fig:escape_lisa}.

\subsection{State of the Art}

The idea of injecting controlled aberration maps to create diversity in the focal plane and use the output to improve the post-processing performance is not new. Both from space and from the ground there have been several attempts to implement this, however, previous developments were tested or implemented mostly for lower order modes (tip-tilt only), or not tested on sky, and in all case not demonstrated at $10^{-8}$ contrast level in a space environment. We review below related techniques published in the literature.

The first application of this idea on space-based facilities is with Hubble Space Telescope (HST), 
where a sub-pixel tip-tilt dithering pattern was used for STIS \cite{Medallon2023_stis} coronagraphy during observations of the reference star. It consists in dithering the reference star in order to simulate misalignments of the star itself behind the coronagraph due to pointing errors. This method successfully improved the starlight subtraction at post-processing stage \cite{debes2019}.
The same approach is in use on the James Webb Space Telescope (JWST \cite{Rigby_2023}), where there is a first actual use of systematic mirror control on space, using the Fine Steering Mirror (FSM \cite{rigby2023_FSM}). These dithered reference images have proven to significantly improve the contrast performance of one order of magnitude \cite{Soummer2014,Lajoie2016}.

Both on HST on JWST the small-grid dithers became a common practice and it is constantly used in the observational routines of these telescopes.

Further studies are carried out on the High-contrast imager for Complex Aperture Telescopes (HiCAT \cite{Soummer2018}) at the Space Telescope Science Institute (STScI). Here they are testing dark hole maintenance algorithms applicable to space observations by using DMs ditherings which help in compensating for the drift of the electric field in the dark hole \cite{Redmond2020}.


From the ground, where wavefront control systems and DMs are common in high contrast imaging instruments, initial developments have been done to investigate using the telemetry of the sensors to improve data post-processing\cite{vogt2011, Singh2016}. Unfortunately, this concept was only preliminary tested through numerical simulations.
In particular, at Subaru - SCExAO \cite{Martinache2009} they inject Zernike error maps on the DMs and create a library that is then used to subtract the coronagraph leakage \cite{vogt2011,Singh2016}. 
A similar work has been carried out at VLT - SPHERE \cite{Beuzit2019}. By using the estimated electric field after the probes, they computed the reference star image model and they subtracted it from the science target. This demonstrated the Classical Differential Imaging technique (CDI \cite{Marois_2006}) on sky. 

\section{SIMULATIONS}
CAPyBARA (Coronagraph and Aberration Python Based Algorithm for Roman Analysis) is a three-level simulator based on \texttt{HCIPy} \cite{por2018hcipy}. Each level is complementary to run a typical Roman observing sequence. Those are: the optical propagation, the environment and the post-processing. 

\subsection{Optical Propagation}

The first level of the simulator consist in the Coronagraph Instrument itself, including both the hardware and the dark hole digging technique.

\subsubsection{Hardware}
\begin{figure} [ht]
\begin{center}
\begin{tabular}{c} 
\includegraphics[height=6.45cm]{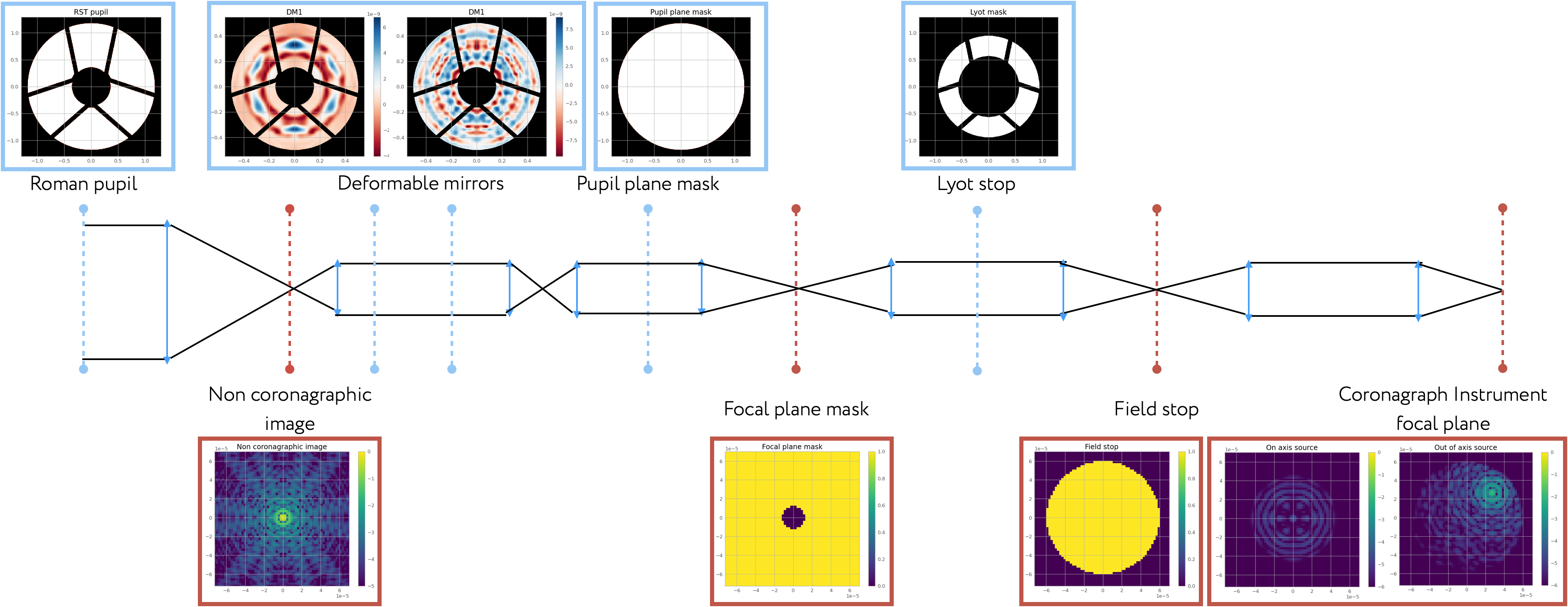}
\end{tabular}
\end{center}
\caption[example] { \label{fig:HLC_optical_path} Optical path of the HLC. On top, highlighted in blue, there are the pupil plane elements, on the bottom, highlighted in red, the focal plane components or images. From left to right we have the Roman entrance pupil, the first non-coronagraphic image, the two DMs, the pupil plane mask (where the shape pupil is inserted in the shaped pupil configuration), the focal plane mask, the Lyot stop mask, the field stop and the last Coronagraph Instrument focal plane (on-axis source on the left and out-of-axis source on the right). The focal plane mask is a simplified version of the original design.}
\end{figure}

The simplified HLC optical path is composed by 6 optical elements (see figure \ref{fig:HLC_optical_path}). Those are, in order, the entrance pupil, the 2 DMs, the focal plane mask, the Lyot stop, the field stop and the final detector. Some of the optical elements \cite{Riggs_2021} that are present in the original design have not been included since they are not relevant for the HLC mode. In this first version of the pipeline, we chose to simulate the focal plane mask with a simple amplitude occulter instead of a phase-amplitude occulter \cite{Byoung_Joon2016}, to simplify the simulations. This relaxation of the original design is not a problem, since our goal is to use the simulator as a proof of concept for new post-processing techniques, and not as the official simulator.


\subsubsection{Dark hole digging}
\begin{figure} [ht]
\begin{center}
\begin{tabular}{c} 
\includegraphics[height=7cm]{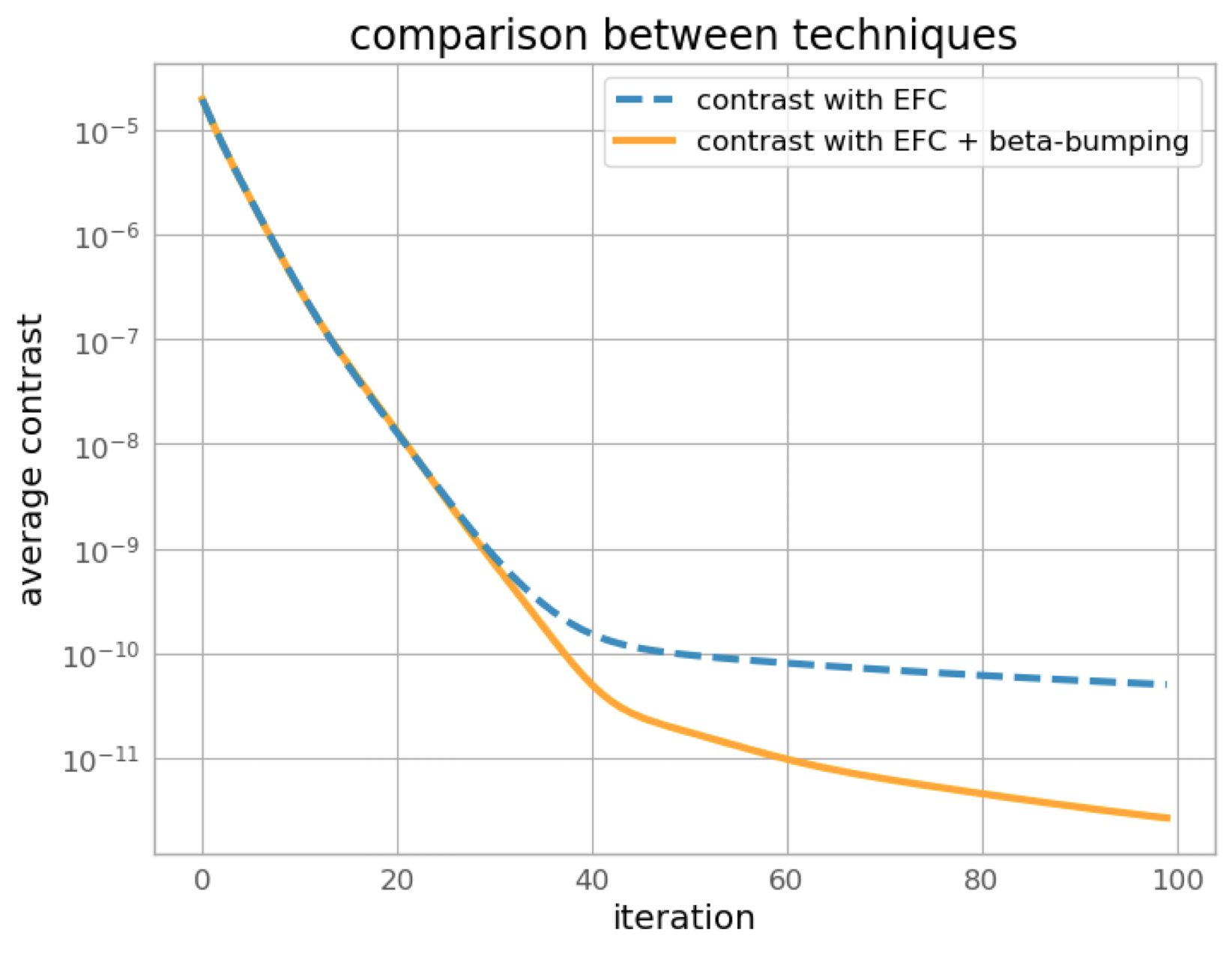}
\end{tabular}
\end{center}
\caption[example] { \label{fig:EFC_beta_bumping} Average contrast in the dark hole area calculated by using only the EFC method (blue dashed line) or EFC method combined with the $\beta$-pumping technique (solid orange line) applied in one iteration every ten for six times. In both cases the simulation does not take into account any aberration. In the second case, the performance is more than one order of magnitude better.}
\end{figure}

In Roman, the dark hole is dug with the Electric Field Conjugation (hereafter EFC) method. It consists in measuring the electric field in the detector and reshape the DMs accordingly to destructively interfere with the injected aberration such that the speckles in the post-coronagraphic focal plane are minimised in a high-contrast area of the field of view between $6 \ \lambda \ D$ and $9 \ \lambda / D$ \cite{giveon_2007}. The electric field is estimated by Pairwise Probing (PWP): a set of linear independent commands is sent to the DMs and the resulting focal plane images are combined to estimate the electric field in each pixel of the dark zone \cite{Krist2023_the_bible}. The estimate of the electric field is then combined with an optical model of the coronagraph, through its Jacobian matrix, to compute the DMs commands that minimize the energy in the dark zone.
This method is coupled with the $\beta$-bumping technique. During the dark hole digging process, only the modes below a certain spatial frequency are suppressed, with the risk of reaching a local minimum. The $\beta$-bumping consists in probing (typically every $10$-$15$ iterations) higher spatial frequencies to redistribute the residuals to the lower frequency that are regularly optimised \cite{Krist2023_the_bible}.
Figure \ref{fig:EFC_beta_bumping} shows a comparison in dark hole digging performance between EFC only and EFC combined with $\beta$-bumping. The difference after $100$ iterations is of more than $1$ order of magnitude, proving that coupling the two techniques helps in reaching higher contrast limits. This simulation is run in a perfect environment where no aberration is added to the system.

\subsection{Environment}
\begin{figure} [ht]
\begin{center}
\begin{tabular}{c} 
\includegraphics[height=7cm]{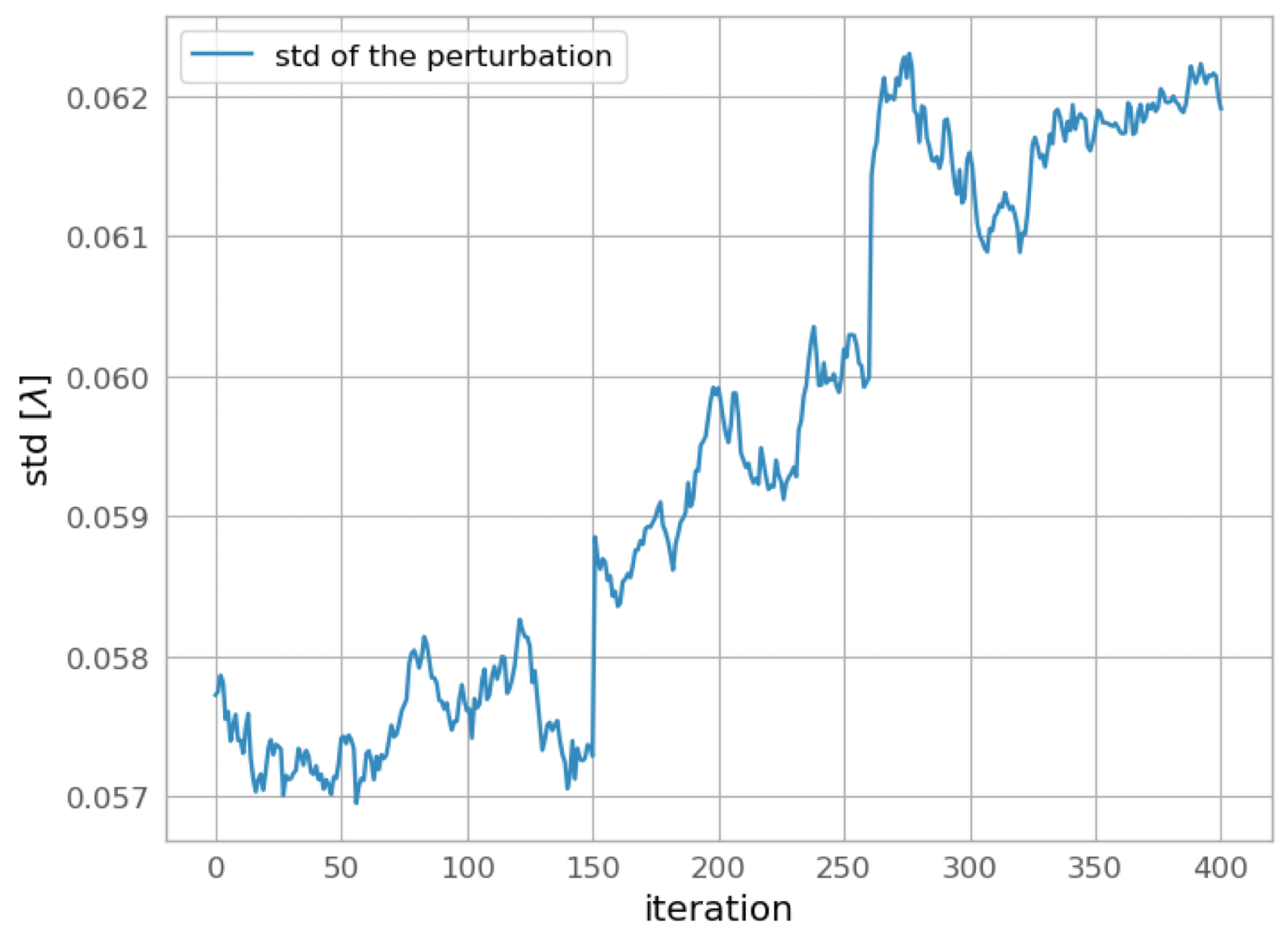}
\end{tabular}
\end{center}
\caption[example] { \label{fig:std_perturbation} Standard deviation of the quasi-static aberration injected in the simulator. The offset at iteration $i=150$ corresponds to the slew of the telescope, the one at iteration $i=260$ corresponds to the roll between one scientific target data acquisition and the following one. The aberration has been created by perturbing a combination of 21 Zernike modes using the formulas presented in Section 2.2.}
\end{figure}

The environment is represented by the quasi-static aberrations that are introduced in the system by thermal fluctuations caused by different insulation and physical movements of the telescope. We started producing a random surface aberration with 21 Zernike modes excluding the first $3$ (piston, tip and tilt), since for those there is a dedicated control mirror, which runs at $20 \ \texttt{Hz}$ \cite{Bailey2023}. Each mode is perturbed independently by using the formulas introduced in Pogorelyuk et al. 2019 \cite{Pogorelyuk2019}, where we added an $\alpha$ parameter that varies between $1$ and $50$ to tune the amplitude of the variation:

\

\begin{gather*} \label{eq1}
    z_p^{p-2j}(i) = 0.99 \cdot z_p^{p-2j} (i-1) + \alpha \cdot \Delta z_p^{p-2j}(i) \\
    \Delta z_p^{p-2j} \sim \mathcal{N} \left( 0, \frac{0.1 \ \texttt{nm}}{p^2 \cdot \lambda} \right) \\
    0 \leq j \leq p \\
\end{gather*}

\noindent where $z_p^{p-2j}(i)$ is the Zernike mode at iteration $i$, $p$ the polynomial order, $p-2j$ the azimuthal degree, $\mathcal{N}$ the normal distribution and $\lambda$ the wavelength.
Each mode is perturbed independently, and then we added them back together to reconstruct the $i$-th aberration. We iterated the process $400$ times and we added a static wavefront error map to simulate the wavefront drift caused by thermal relaxation after slewing from a reference star to the scientific target ($i=150$) and the roll of the telescope to introduce angular diversity during the observations of the science target ($i=260$), which is the default observing strategy for the Roman Coronagraph observations. The offset is randomly generated in the same way as the surface aberration at $i=0$.
The evolution of the standard deviation during the process is plotted in Figure \ref{fig:std_perturbation}, where we used a tuning parameter $\alpha =50$. 

\subsection{Post-processing}
A typical Roman observing sequence is divided into three main steps (Figure \ref{fig:contrast_observing_sequence}): dark hole digging, reference star acquisition and science target acquisition, sub-divided into rolls. Between one roll and the other, the telescope rotates of $\pm 13^{\circ}$. Then, the data from the reference star and the science target are combined and used during the post-processing phase.

For Roman data, several post-processing techniques have been applied on the Observing Scenarios (OS \footnote{\url{https://roman.ipac.caltech.edu/sims/Coronagraph_public_images.html}}) released by IPAC. Currently, those that have been explored are: classical Reference-star Differential Imaging (cRDI), classical Angular Differential Imaging (cADI) and KL image projection RDI (KLIP RDI \cite{Soummer2012}), which is a Principle Component Analysis (PCA) technique.
Despite there is a different performance depending on the case (noisy or noiseless dataset), all the techniques perform better than the design requirement of $5 \times 10^{-8}$ in flux ratio sensitivity \cite{Ygouf2021}.
In this work, we will present a first post-processing result by applying cRDI (Section 3.2).

\section{RESULTS}
\label{sec:sections}
The pipeline described in Section 2 allows us to simulate a typical observing sequence of the Roman Coronagraph and produce post-processing science images. As a first simulation, we run a Roman Coronagraph Instrument simplified observing sequence. It differs from the designed one because it is shorter: each reference star data acquisition is followed by two science rolls (A-B) instead of four (A-B-A-B).

\subsection{Data Acquisition}
\begin{figure} [ht]
\begin{center}
\begin{tabular}{c} 
\includegraphics[height=8cm]{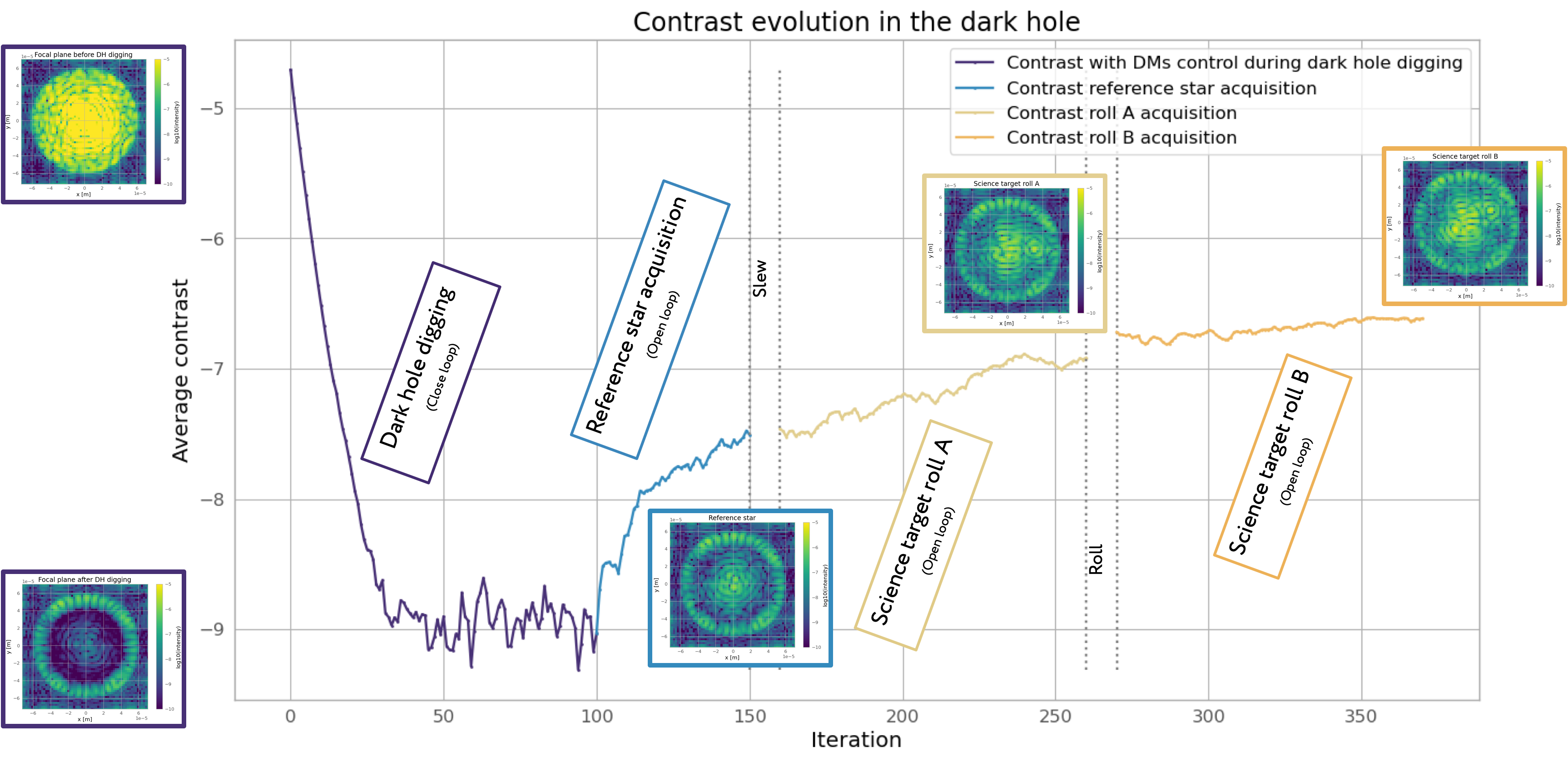}
\end{tabular}
\end{center}
\caption[example] { \label{fig:contrast_observing_sequence} Average contrast in the dark hole area during each stage of a typical Roman observing sequence. Each of the framed images shows the final focal plane, from left to right, at the beginning (top) and at the end (bottom) of the dark hole digging process, at the end of the reference star acquisition, at the end of the scientific target data acquisition in roll A and in roll B. The $\alpha$ parameter in the quasi-static aberration simulator was set equal to $50$.}
\end{figure}

We start with a post-coronagraph image where the focal plane is dominated by speckles and the average contrast is of the order of $10^{-5}$ (Figure \ref{fig:contrast_observing_sequence}, top frame on the left). During the dark hole digging, CAPyBARA runs in closed loop. The DMs are re-shaped to compensate for the quasi-static aberration until we reached a plateau corresponding to a contrast ratio of $10^{-9}$ (Figure \ref{fig:contrast_observing_sequence}, bottom frame on the left). Once this limit is stable, we perform the reference star data acquisition in open loop, without compensating for the quasi-static aberration (Figure \ref{fig:contrast_observing_sequence}, third frame from the left). From the plot we can see that the contrast limit immediately starts to downgrade as soon as the aberration is not anymore corrected. In the end, after $50$ iterations, we lose $1.5$ order of magnitudes in contrast. We then simulate a slew to the science target and we performed two data acquisition. Between roll A (Figure \ref{fig:contrast_observing_sequence}, fourth frame from the left) and roll B (Figure \ref{fig:contrast_observing_sequence}, last frame from the left), the telescope is rotated of $26^{\circ}$. Overall, during the science target acquisition, we lost about one order of magnitude in contrast limit. This value is strictly dependent on how fast the quasi-static aberration evolves and, thus, on which value for the $\alpha$ parameter we choose. 

\subsection{Post-Processing Results}
\begin{figure} [ht]
\begin{center}
\begin{tabular}{c} 
\includegraphics[height=5cm]{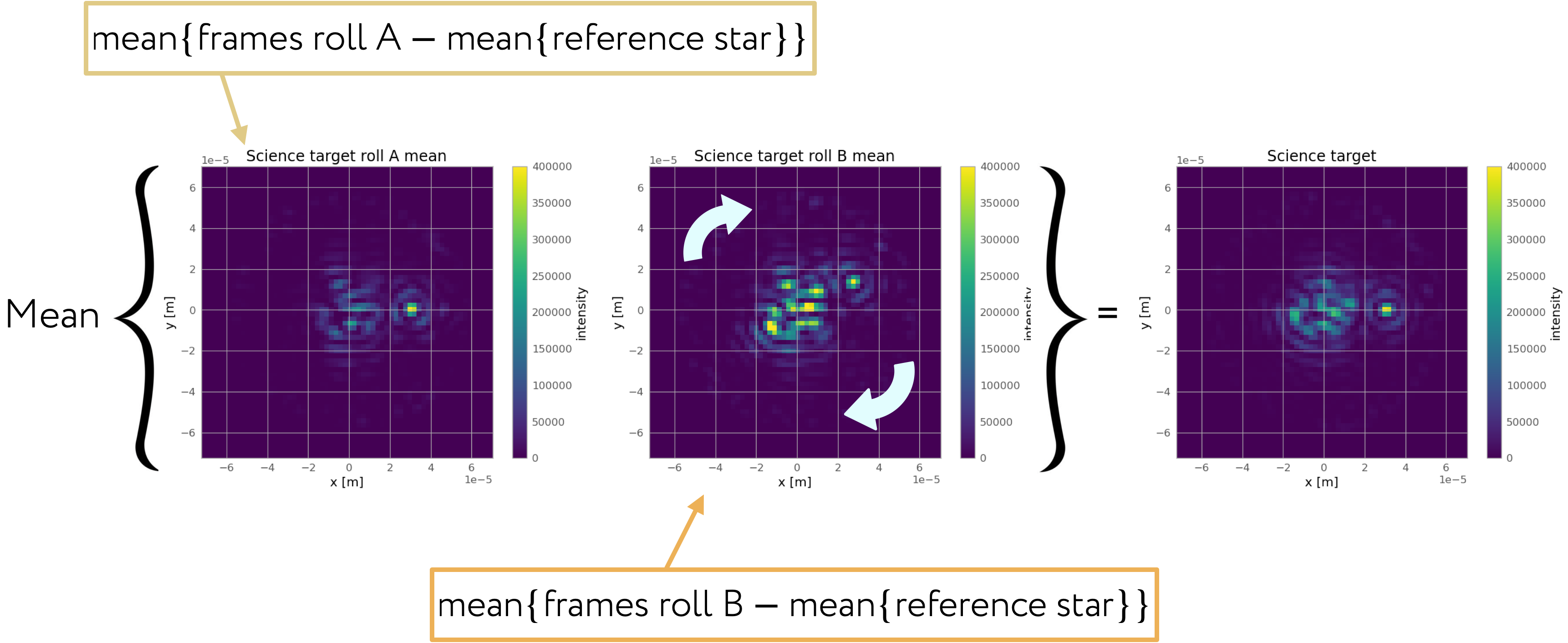}
\end{tabular}
\end{center}
\caption[example] { \label{fig:post_processing} Schematic summary of the post-processing analysis applied to the simulated dataset by using classical RDI.}
\end{figure}

As anticipated in the Section 2.3, here we will present a first post-processing result by applying RDI, one of the classical PSF subtraction techniques.
From the data acquisition, we have three datasets: one for the reference star and two for the scientific target, one for each roll.
We started by finding the mean of all the frames of the reference star and we subtracted it to each frame of the scientific target. Then, we computed the average image of the scientific target in each of the two rolls, which for simplicity we will call \emph{mean roll A} and \emph{mean roll B}. We derotated \emph{mean roll B} by $26^{\circ}$ and we calculated the mean between the derotated \emph{mean roll B} and \emph{mean roll A}. Figure \ref{fig:post_processing} summarizes the procedure.
Since the dataset simulated with CAPyBARA are in monochromatic wavelength, post-processing results are not fully realistic and representative of a typical Roman dataset. For this reason, in this section we will focus more on the qualitative analysis of the final result. 
By comparing \emph{mean roll A} with \emph{mean roll B}, there is an evident difference in the average residuals from the subtraction of the PSF reference star, and the intensity of the planet is comparable to that of the speckles. This is expected, since the average contrast deteriorates of almost one order of magnitude from the first to the last frame of the scientific target dataset. Despite this, in the final image, obtained by averaging the two rolls, the contribution from the speckles attenuates of almost a factor of $2$. Although, we would like to point out that the presence of the planet was already inferable from the raw images. 
Overall, the post-processing method improved by far the contrast limit around the star. This result, obtained with a classical PSF subtraction, sets a good starting point that makes us believe in much better results with more advanced post-processing techniques. 
Although, we should keep in mind that these data were simulated in monochromatic wavelength, so we expect to have a loss in performance once implementing the broadband. 

\section{CONCLUSION}
In this paper we presented a first version of CAPyBARA, which simulates the Coronagraph Instrument from the optical propagation, the environment where it will operate and the post-processing analysis. With it, we run a simplified version of a typical Roman observing sequence. We showed a first result by injecting a synthetic planet and by analysing the data with a classical PSF subtraction (RDI).
Currently, the the main limitation of the simulator is that it runs in monochromatic wavelength. In order to make the data more realistic, the next step will be to extend the simulator to broadband.
In addition to this, in CAPyBARA we will implement the PCA analysis among the post-processing techniques, as well as the dark hole touch up, that in the nominal observing sequence occurs after two reference+scientific target data acquisitions. Once this is implemented, we will have a more realistic simulator and with it we will run longer observing sequences.
Last, we will investigate the calibrated high-order mode dithering by injecting controlled calibration maps on the DMs, which is the final goal of the project.

\acknowledgments      
   
This project is funded by the European Union (ERC, ESCAPE, project No 101044152). Views and opinions expressed are however those of the authors only and do not necessarily reflect those of the European Union or the European Research Council Executive Agency. Neither the European Union nor the granting authority can be held responsible for them.  


\bibliographystyle{spiebib} 

\end{document}